\documentclass[a4paper]{article}
\usepackage{amsmath,amssymb}
\usepackage{graphicx}

\def\be {\begin{equation}}
\def\ee {\end{equation}}
\def\bea {\begin{eqnarray}}
\def\eea {\end{eqnarray}}
\newcommand{\nn}{\nonumber \\}
\newcommand{\e}{\mathrm{e}}

\DeclareMathOperator{\sech}{sech}

\begin{document}

\title{Generalized $f(R,\phi,X)$ gravity and the late-time cosmic acceleration}

\author{
Sebastian Bahamonde$^{1}$\footnote{sebastian.beltran.14@ucl.ac.uk} , 
Christian G. B\"ohmer$^{1}$\footnote{c.boehmer@ucl.ac.uk} , \\ 
Francisco S.N. Lobo$^{2,3}$\footnote{fslobo@fc.ul.pt} , 
and Diego S\'{a}ez-G\'{o}mez$^{2}$\footnote{dsgomez@fc.ul.pt}\\[1ex]
$^{1}$ Department of Mathematics, University College London,\\ 
Gower Street, London, WC1E 6BT, UK \\[0.5ex]
$^{2}$ Instituto de Astrof\'{\i}sica e Ci\^{e}ncias do Espa\c{c}o, Universidade de Lisboa, \\
Campo Grande, PT1749-016 Lisboa, Portugal\\[0.5ex]
$^{3}$ Faculdade de Ci\^encias da Universidade de Lisboa, \\ 
Campo Grande, PT1749-016 Lisboa, Portugal
}

\date{7 Aug 2015}

\maketitle

\begin{abstract}
High-precision observational data have confirmed with startling evidence that the Universe is currently undergoing a phase of accelerated expansion. This phase, one of the most important and challenging current problems in cosmology, represents a new imbalance in the governing gravitational equations. Historically, physics has addressed such imbalances by either identifying sources that were previously unaccounted for, or by altering the gravitational theory. Several candidates, responsible for this expansion, have been proposed in the literature, in particular, dark energy models and modified gravity models, amongst others. Outstanding questions are related to the nature of this so-called ``dark energy'' that is driving this acceleration, and whether it is due to the vacuum energy or a dynamical field. On the other hand, the late-time cosmic acceleration may be due to modifications of General Relativity. In this work we explore a generalised modified gravity theory, namely $f(R,\phi,X)$ gravity, where $R$ is the Ricci scalar, $\phi$ is a scalar field, and $X$ is a kinetic term. This theory contains a wide range of dark energy and modified gravity models. We considered specific models and applications to the late-time cosmic acceleration.
\end{abstract}

\clearpage

\section{Introduction}

A central theme in Cosmology is the fact that the Universe is currently undergoing an accelerating expansion \cite{Perlmutter:1998np,Riess:1998cb}. In this context, during the last two decades Cosmology has evolved from being mainly a theoretical area of Physics to become a field supported by observational data. Recent experiments call upon state of the art technology in Astronomy and Astrophysics to provide detailed information on the contents and history of the Universe, which has led to the measuring of the parameters that describe our Universe with increasing precision. The standard model of cosmology is remarkably successful in accounting for the observed features of the Universe. However, there remain a number of open questions concerning the foundations of that standard model. In particular, we lack a fundamental understanding of the mechanisms underlying the acceleration of the late-time universe. What is the so-called ``dark energy'' that is driving the cosmic acceleration? Is it a vacuum energy or a dynamical field? Is the acceleration due to modifications of Einstein's theory of General Relativity (GR)? How is structure formation affected in these alternative scenarios? What happens to the universe in the asymptotic future?

The resolution of these fundamental questions, looking beyond the standard theories of gravity and particle physics, is extremely important for theoretical cosmology and theoretical physics as a whole. The standard model of cosmology has favoured dark energy models as fundamental candidates responsible for the cosmic expansion. However, it is clear that these questions involve not only gravity, but also particle physics. String theory provides a synthesis of these two branches of physics and is widely believed to be moving towards a viable quantum gravity theory, given time. One of the key predictions of string theory is the existence of extra spatial dimensions. In the brane-world scenario, motivated by recent developments in string theory, the observed 4-dimensional universe is embedded in a higher-dimensional spacetime \cite{Maartens:2003tw}. The new degrees of freedom belong to the gravitational sector, and can be responsible for the late-time cosmic acceleration \cite{Dvali:2000hr,deRham:2007rw}. On the other hand, generalisations of the Einstein-Hilbert Lagrangian, including quadratic Lagrangians which involve second order curvature invariants have also been extensively explored \cite{Sotiriou:2008rp,DeFelice:2010aj,Capozziello:2011et,Lobo:2008sg,Nojiri:2010wj,delaCruzDombriz:2012xy}. An alternative approach was formulated in \cite{Boehmer:2013ss,Boehmer:2014ipa}. 
Curvature-matter couplings have also been extensively analysed \cite{Allemandi:2005qs,Bertolami:2007gv,Harko:2010hw,Harko:2014gwa,Harko:2011kv,Haghani:2013oma,Odintsov:2013iba}. While these modified theories of gravity offer an alternative explanation to the standard cosmological model for the expansion history of the universe \cite{Carroll:2003wy}, it offers a paradigm for nature fundamentally distinct from dark energy models of cosmic acceleration \cite{Copeland:2006wr}, even those that perfectly mimic the same expansion history. It is also fundamental to understand how one may differentiate these modified theories of gravity from dark energy models.

In this context, a promising way to explain the late-time cosmic acceleration is to assume that at large scales Einstein's theory of GR breaks down, and a more general action describes the gravitational field. Thus, one may generalise the Einstein-Hilbert action by including ``quadratic Lagrangians'', involving second order curvature invariants. Some of the physical motivations for these modifications of gravity were related to the possibility of a more realistic representation of the gravitational fields near curvature singularities and to create some first order approximation for the quantum theory of gravitational fields. On the other hand, field theories in theoretical physics tends to have Lagrangians quadratic in field strengths, gravity being somewhat different. One may tackle the problem using the metric formalism, which consists in varying the action with respect to the metric, although other alternative approaches have been considered in the literature, namely, the Palatini formalism \cite{Olmo:2011uz}, where the metric and the connections are treated as separate variables; and the metric-affine formalism, where the matter part of the action now depends and is varied with respect to the connection \cite{Sotiriou:2006qn}. Recently, a novel approach to modified theories of gravity that consists of adding to the Einstein-Hilbert Lagrangian an $f(R)$ term constructed a la Palatini \cite{Harko:2011nh}. It was shown that the theory can pass the Solar System observational constraints even if the scalar field is very light. This implies the existence of a long-range scalar field, which is able to modify the cosmological and galactic dynamics, but leaves the Solar System unaffected. These explicit models are consistent with local tests and lead to the late-time cosmic acceleration, and also verify the absence of instabilities in perturbations. 

In this work, we intend to consider a generalised modified gravity theory, $f(R,\phi,X)$ gravity, where $R$ is the Ricci scalar, $\phi$ a scalar field, and $X$ a kinetic term. This theory contains a wide range of known dark energy and modified gravity models, for instance $f(R)$ gravity models or Galileons. Here we apply it to the late-time cosmic acceleration. In particular, several cosmological solutions are studied within the framework of these theories, specifically solutions that can provide cosmic acceleration at late times, and even the exact $\Lambda$CDM evolution. Reconstructions techniques are implemented in order to obtain the $f(R,\phi,X)$ of the action given a particular Hubble parameter. This provides a way to efficiently check the viability of any gravitational action by just considering a particular cosmological evolution and then analysing the gravitational action. An action of the form $f(R,\phi,X)$ is in fact quite natural as it removes any assumptions on the underlying theory of gravity with the exception of being second order. We can think of the field $\phi$ as the effective field controlling the strength of the gravitational force. One should emphasise that normal matter is still coupled minimally to this theory.

This paper is organised in the following manner: In Section \ref{sec2}, we present the general formalism of $f(R,\phi,X)$ gravity. In Section \ref{sec3}, we analyse cosmological applications, in particular, to the late-time cosmic acceleration, by considering specific models and reconstructing the corresponding gravitational action. Then, we discuss in the conclusions at section \ref{conclusions} about the results obtained and possible future considerations of these theories.


\section{Generalised gravity models: formalism}\label{sec2}

Consider the following general action
\begin{equation}
  S=\int d^4x \sqrt{-g}\left[ \frac{1}{2\kappa^2} f(R,\phi,X) + L_m \right] \,,
  \label{action}
\end{equation}
where $f=f(R,\phi,X)$ is a function of the Ricci scalar $R$, the scalar field $\phi$, and a kinetic term $X=-(1/2)(\nabla \phi)^2$. The matter Lagrangian is $L_m=L_m(g, \psi)$, we note that matter is minimally coupled to the gravitational sector of the theory and $\psi$ collectively denotes any matter fields.

The variation of action (\ref{action}) with respect to the metric $g_{\mu\nu}$ gives to the following field equation
\begin{equation}
  F G_{\mu\nu}=\frac{1}{2}\left( f-RF  \right) g_{\mu\nu} + \nabla_\nu F_{,\mu}-
  g_{\mu \nu}\nabla_\alpha \nabla^\alpha F +\frac{1}{2}f_{,X} \, \phi_{,\mu}\phi_{,\nu} + T_{\mu\nu}^{({\rm m})} \,,
   \label{fieldeq1}
\end{equation}
while variations with respect to the scalar field $\phi$ yield
\begin{equation}
  \nabla_\mu \left( f_{,X}\, \phi^{,\mu}  \right) + f_{,\phi}=0 \,.
  \label{fieldeq2}
\end{equation}
Here $F=\partial f/\partial R$ and $T_{\mu\nu}^{({\rm m})}$ is the matter energy-momentum tensor defined by
\begin{equation}
  T_{\mu\nu}^{({\rm m})} = - \frac{2}{\sqrt{-g}}\frac{\delta(\sqrt{-g}\,L_m)}{\delta(g^{\mu\nu})} \,.
\end{equation}    

Now, taking into account a flat Friedman-Lema\^{\i}tre-Robertson-Walker (FLRW) metric in spherical coordinates
\begin{equation}
  ds^2=-dt^2+a^2(t)\left[dr^2+r^2\left(d \theta^2+\sin^2\theta d\varphi\right)\right]\,,
  \label{FLRW}
\end{equation}
where $a(t)$ is the scale factor, then Eqs.~(\ref{fieldeq1}) and (\ref{fieldeq2}) take the following form
\begin{eqnarray}
  3FH^2 &=& f_{,X}X + \frac{1}{2}\left( FR-f \right) - 3H \dot{F} + \kappa^2 \rho_{\rm m} \,,
  \label{modFriedeq1}  \\
  -2F\dot{H} &=& f_{,X}X + \ddot{F} - H \dot{F} +  \kappa^2 (\rho_{\rm m}+p_{\rm m}) \,,
  \label{modFriedeq2} \\
  0 & = & \frac{1}{a^3}\left( a^3 \dot{\phi} f_{,X} \right)^{.} - f_{,\phi} \,,
\end{eqnarray}
where $H=\dot{a}/a$ is the Hubble parameter and the overdot denotes a derivative with respect to cosmological time $t$. For simplicity we will not take into account additional forms of matter.

In order to derive an equation for the effective equation of state, we rewrite the field equations (\ref{modFriedeq1})--(\ref{modFriedeq2}) in the following form
\begin{eqnarray}
  3H^2 &=& \kappa^2 \left( \rho_{\rm m} + \rho_{\rm DE} \right) \,,
  \label{modFriedeq1b} \\
  -2\dot{H} &=& \kappa^2 \left( \rho_{\rm m} + p_{\rm m} + \rho_{\rm DE} +p_{\rm DE} \right)\,, 
  \label{modFriedeq2b}
\end{eqnarray}
where the quantities $\rho_{\rm DE}$ and $p_{\rm DE}$ are defined by
\begin{eqnarray}
  \kappa^2 \rho_{\rm DE}&=& f_{,X}X + \frac{1}{2}\left( FR-f \right) - 3H \dot{F} +3H^2(1-F)\,,
  \label{defDE1} \\
  \kappa^2 p_{\rm DE}&=& \ddot{F} + 2H\dot{F} - \frac{1}{2}
  \left( FR-f \right) - \left(2\dot{H} +3H^2 \right)(1-F) \,.
  \label{defDE2}
\end{eqnarray}
One may easily show that the these effective dark energy components satisfy the usual conservation equation, i.e., 
\begin{equation}
\dot{\rho}_{\rm DE}+3H\left( \rho_{\rm DE}+p_{\rm DE} \right)=0\,.
\end{equation} 
This is expected since the left-hand sides of (\ref{modFriedeq1b})--(\ref{modFriedeq2b}) are equivalent to the general relativistic field equations containing the Einstein tensor which satisfies the twice contracted Bianchi identities. 

Let us define the dark energy equation of state given by $\omega_{\rm DE} \equiv  p_{\rm DE}/\rho_{\rm DE}$, which by using Eqs.~(\ref{modFriedeq1b})--(\ref{modFriedeq2b}) takes the form
\begin{equation}
  \omega_{\rm DE} = - \frac{2\dot{H} +3H^2 + \kappa^2 p_{\rm m}}{3H^2 - \kappa^2 \rho_{\rm m}} \,.
\end{equation}
In order to specify the matter content, let us choose the equation of state $p_{\rm m} = w \rho_{\rm m}$. In turn, we can write the dark energy equation of state as
\begin{equation}
  \omega_{\rm DE} = - \frac{1 + \frac{2}{3}\frac{\dot{H}}{H^2} + w\frac{\kappa^2 \rho_{\rm m}}{3H^2}}{1-\frac{\kappa^2 \rho_{\rm m}}{3H^2}} = \frac{\omega_{\rm eff}}{1- \tilde{\Omega}_m} \,,
\end{equation}
where 
\begin{equation}
\omega_{\rm eff}= -1 - \frac{2}{3}\frac{\dot{H}}{H^2} -w\tilde{\Omega}_m, \qquad  \tilde{\Omega}_m = \frac{\kappa^2 \rho_m}{3H^2} = F\,\Omega_m \,.
\end{equation}

In the following we will consider the above equations and seek solution corresponding to an accelerated expansion.

\section{Cosmological applications: Late-time cosmic acceleration}
\label{sec3}

Let us consider some specific forms of the action (\ref{action}) and reconstruct particular cosmological solutions capable of reproducing the late-time cosmic acceleration. We are considering pure de Sitter solutions, also power law solutions and finally we are reconstructing an exact $\Lambda$CDM model. Note that this kind of cosmological solutions have already been considered in the literature within modified gravity theories, specially in $f(R)$ gravity and Gauss-Bonnet gravity. In particular, power-law solutions were found in \cite{Goheer:2009ss,Myrzakulov:2010gt} while exact $\Lambda$CDM model were reconstructed in \cite{delaCruzDombriz:2006fj,Elizalde:2010jx,Dunsby:2010wg}. In order to simplify the calculations, we restrict to a matter Lagrangian with a pressure-less fluid, as natural when analysing late-time cosmology. 

\subsection{Brans-Dicke type models}

Here we consider a Brans-Dicke type action with $f(R,X,\phi)=\gamma(X,\phi)R$, where the coupling to the Ricci curvature also includes the kinetic term of the scalar field,
\begin{equation}
  S = \int d^4x \sqrt{-g}\left[\frac{1}{2\kappa^2}\gamma(X,\phi)R + L_{\rm m}\right] \,.
\label{action1}
\end{equation}
Then, by considering an homogeneous and isotropic metric (\ref{FLRW}), the FLRW equations (\ref{modFriedeq1}, \ref{modFriedeq2}) are expressed as follows:
\begin{eqnarray}
  3\gamma(X,\phi)H^2&=&\gamma_{,X}X R-3H\frac{d}{d t}\gamma(X(t),\phi(t))+
  \kappa^2\rho_m \,,
  \nn
  -2\gamma(X,\phi)\dot{H}&=&\gamma_{,X}X R+\frac{d^2}{d t^2}\gamma(X(t),\phi(t))-
  H\frac{d}{d t}\gamma(X(t),\phi(t))+\kappa^2(\rho_{\rm m}+p_{\rm m}) \,.
  \label{FLRW1}
\end{eqnarray}
And combining both equations, a differential equation solely of $\gamma(X(t),\phi(t))=\gamma(t)$  is obtained
\begin{equation}
  \frac{d^2 \gamma(t)}{d t^2}+2H\frac{d \gamma(t)}{dt}+(3H^2+2\dot{H})\gamma(t) = 0 \,.
\label{Comb}
\end{equation}
Then, equation (\ref{Comb}) can be solved by assuming a particular Hubble parameter and action (\ref{action1}) can be reconstructed by assuming a particular form of $\gamma(X,\phi)$, which has to satisfy
\begin{equation}
  \gamma_{,X}X=\frac{3\gamma H^2+3H\dot{\gamma}-\kappa^2\rho_{\rm m}}{R} \,,
  \label{D1}
\end{equation}

As an example for illustrating the reconstruction method, here we consider the usual form of a scalar field Lagrangian,
\begin{equation}
\gamma(X,\phi)=X-V(\phi) \,,
\label{D2}
\end{equation}
where $V(\phi)$ is a function of the scalar field to be determined. Then, by redefining the scalar field to coincide with the cosmic time $\phi=t$ and using (\ref{D1}) and (\ref{D2}), the kinetic term and the scalar potential are fully reconstructed:
\begin{eqnarray}
X(\phi)&=&\frac{3\gamma H^2+3H\dot{\gamma}-\kappa^2\rho_0\e^{-3\int d\phi H}}{R} \,, \nn
V(\phi)&=&X(\phi)-\gamma(\phi) \,.
\label{D3}
\end{eqnarray}
where we have used the continuity equation $\dot{\rho}_{\rm m}+3H\rho_{\rm m}=0$.

Let us now consider several cosmological solutions. Firstly we assume a pure de Sitter solution,
\begin{equation}
  H=H_0 \,,
  \label{ds}
\end{equation}
where $H_0$ is a constant. Then, the equation (\ref{Comb}) leads to
\begin{equation}
  \frac{d^2 \gamma(t)}{dt^2}+2H_0\frac{d\gamma(t)}{dt}+3H_0^2\gamma(t)=0 \,.
  \label{CombdS}
\end{equation}
which is the equation of a damped harmonic oscillator, whose general solution is given by
\begin{equation}
  \gamma(t)=\e^{-H_0 t}C\cos\sqrt{2}H_0(t-t_0) \,,
  \label{SoldS}
\end{equation}
where $\{C, t_0\}$ are integration constants. Then, assuming (\ref{D2}), the gravitational action is reconstructed  by the expressions (\ref{D3}), where the kinetic term and the scalar potential turn out
\begin{eqnarray}
  X(\phi)&=&-\frac{C}{2\sqrt{2}}\e^{-H_0\phi}\sin \sqrt{2}H_0(\phi-\phi_0)-
  \frac{\kappa^2\rho_0}{12H_0^2}\e^{-3H_0\phi} \,, \nn
  V(\phi)&=&-\frac{C}{4}\e^{-H_0\phi} 
  \left[4\cos \sqrt{2}H_0(\phi-\phi_0)+\sqrt{2}\sin \sqrt{2}H_0(\phi-\phi_0)\right]-
  \frac{\kappa^2\rho_0}{12H_0^2}\e^{-3H_0\phi} \,,
\label{D4}
\end{eqnarray}
The interesting point of this action is that leads to a constant Hubble parameter in the presence of dust matter and a scalar field but without a cosmological constant, contrary to standard General Relativity. 

Another important class of solutions in cosmology is the power-law type expansion, whose Hubble parameter is given as follows
\begin{equation}
  H=\frac{n}{t}  \quad \rightarrow \quad a(t)=a_{0} t^n  \,.
\label{D5}
\end{equation}

By applying the same procedure as above, the equation (\ref{Comb}) turns out to be
\begin{equation}
  t^2 \frac{d^2 \gamma(t)}{dt^2}+2nt\frac{d\gamma(t)}{dt}+n(3n-2)\gamma(t)=0 \,,
\label{D6}
\end{equation}
which is an Cauchy-Euler equation whose general solution is given by
\begin{equation}
  \gamma(t)=C_1 t^{p+q}+C_2 t^{p-q} \,,
  \label{D7}
\end{equation}
where $C_1$ and $C_2$ are two constants of integration, and we defined 
\begin{equation}
  p=\frac{1-2n}{2} \,, \qquad q=\frac{\sqrt{1+4n-8n^2}}{2} \,.
\end{equation}

By assuming (\ref{D2}) as above, the  kinetic term and the scalar potential are obtained
\begin{eqnarray}
  X(\phi)&=&\frac{1}{4(2n-1)}\Big[C_{1}(1+2q)\phi^{p+q}+C_{2}(1-2q)\phi^{p-q}-
    \frac{1}{3N}2\kappa^2\rho_{0}\phi^{2-3n}\Big] \,, \nn
  V(\phi)&=&\frac{1}{4(2n-1)}\Big[C_{1}(5-8n+2q)\phi^{p+q}+C_{2}(5-8n-2q)\phi^{p-q}-
    \frac{1}{3N}2\kappa^2\rho_{0}\phi^{2-3n}\Big] \,.   
  \label{D8}
\end{eqnarray}  
Hence, the gravitational action capable of reproducing the power-law solutions (\ref{D5}) is obtained.
 
Finally, let us consider the case of exact $\Lambda$CDM and explore whether a gravitational action of the type (\ref{action1}) is capable of reproducing such kind of expansion. In order to simplify the calculations, here we use the redshift $1+z=\frac{1}{a}$ as the independent variable instead of the cosmic time $t$. The cosmological evolution of  $\Lambda$CDM model can be expressed by the following Hubble parameter
\begin{equation}
  H(z)=H_0\sqrt{\Omega_m(1+z)^3+1-\Omega_m} \,,
  \label{LCDM}
\end{equation}
 where $\Omega_m=\frac{\rho_0}{3H_0^2/\kappa^2}$ and $H_0$ is the Hubble parameter evaluated at $z=0$. Then, the equation (\ref{Comb}) can be rewritten in terms of the redshift as,
\begin{equation}
  (1+z)^2H^2 \gamma^{\prime\prime}+\left[(1+z)^2HH^{\prime}-(1+z)H^2\right]\gamma^{\prime}+\left[3H^2-2(1+z)HH^{\prime}\right]\gamma=0 \,,
 \label{D9}
 \end{equation}
 where the primes denote derivatives with respect to the redshift. Hence, the equation (\ref{D9}) can be solved as $\gamma(X,\phi)=\gamma(z)$. Particularly, for the $\Lambda$CDM model, the equation (\ref{D9}) leads to
\begin{equation}
  2(1+z)^2 \left[\Omega_m z(z^2+3z+3)+1\right] \gamma^{\prime\prime}(z)+(1+z) \left[\Omega_m(z^3+3z^2+3z+3) -2\right] \gamma^{\prime}(z)+6 (1-\Omega_m) \gamma (z)=0 \,.
  \label{D10}
\end{equation}

\begin{figure}
  \begin{center}
    \includegraphics[width=0.45\textwidth]{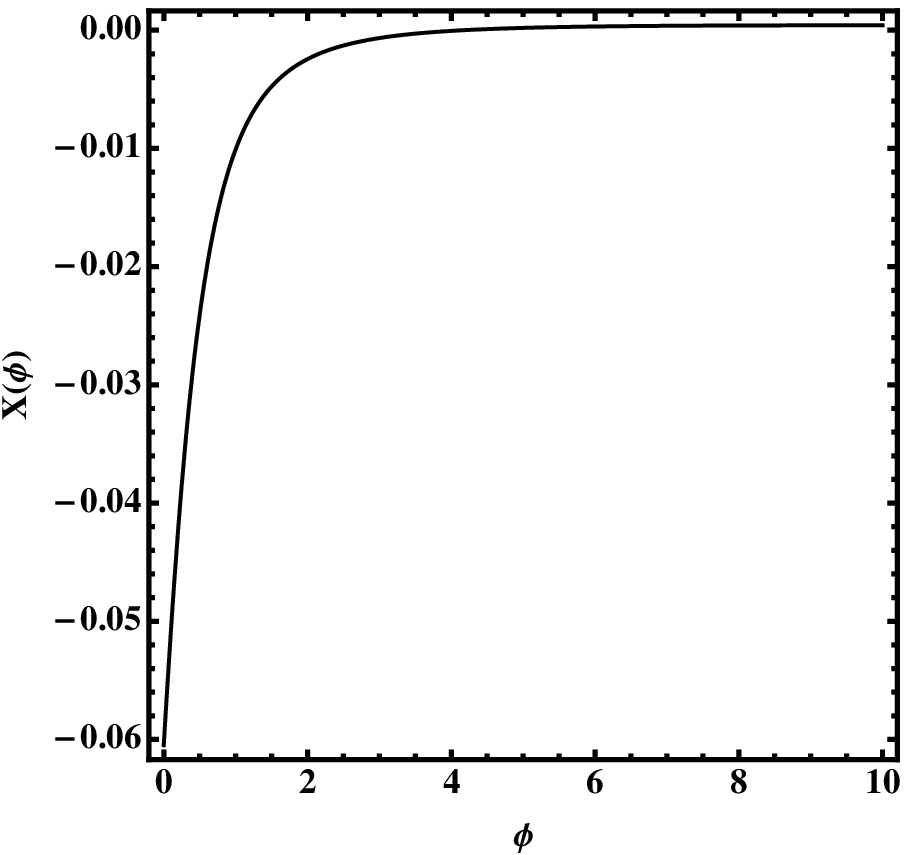}
    \hfill
    \includegraphics[width=0.45\textwidth]{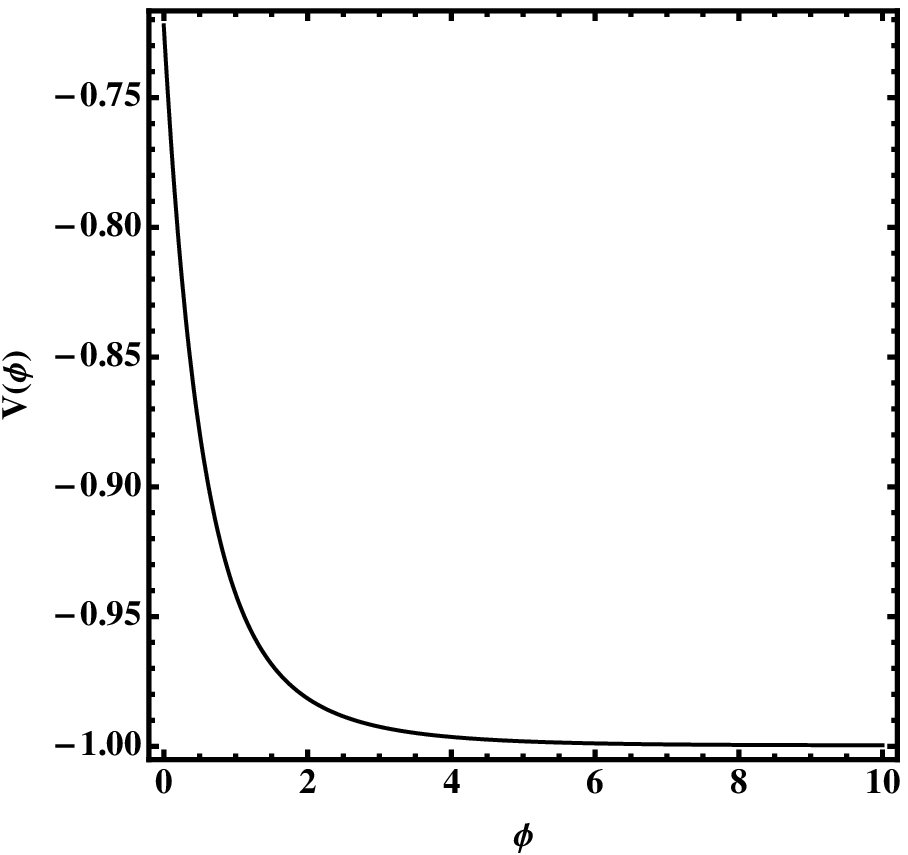}
  \end{center}
  \caption{The kinetic term (left panel) and scalar potential (right panel) for the $\Lambda$CDM model (\ref{LCDM}) with $\Omega_m=0.31$.}
  \label{fig1}
\end{figure}

Unfortunately, this equation can not be solved exactly but numerical resources are required. In order to illustrate a particular case, we solve the equation (\ref{D10}) by assuming $\Omega_m=0.31$, provided by the last Planck data \cite{Ade:2015xua}. The kinetic term and the scalar potential are depicted in Fig.~\ref{fig1}. Note that we have assumed the initial conditions $\gamma(z=10)=1$ and $\gamma'(z=10)=0$, so that GR is recovered at large redshifts. As expected, the dynamics of the scalar field become important at small redshifts while remains virtually constant at large ones.

\subsection{Minimally coupled field model}

Let us now consider another different type of the gravitational action (\ref{action}), where the scalar field is minimally coupled. We choose $f(R,X,\phi)=\alpha(R)+\gamma(X,\phi)$ so that 
\begin{equation}
  S = \int dx^4 \sqrt{-g}\left[\alpha(R)+\gamma(X,\phi) + L_{\rm m} \right] \,.
\label{action2}
\end{equation}
This action can be seen as an $f(R)$-like action with a k-essence-like fluid in the presence of $f(R)$ gravity. Then, the FLRW equations (\ref{modFriedeq1}, \ref{modFriedeq2}) turn out
\begin{eqnarray}
  3\alpha_{,R}H^2&=& \gamma_{,X}X + \frac{1}{2}\left( \alpha_{,R} R-\alpha-\gamma \right) - 
  3H \dot{\alpha}_{,R} + \kappa^2 \rho_m \,, \nn
  -2\alpha_{,R}\dot{H} &=& \gamma_{,X}X + \ddot{\alpha}_{,R} - 
  H \dot{\alpha}_{,R} +  \kappa^2 \rho_m \,,
  \label{D11}
\end{eqnarray}
which can be combined, so that an equation for $\gamma(X,\phi)=\gamma(t)$ is obtained,
\begin{equation}
  \gamma(t)=-\alpha+\left(R-6 H^2-4 \dot{H}\right) \alpha_{,R}-\left(4 H \dot{R}+2 \ddot{R}\right) \alpha_{,RR}-2\dot{R}^2\alpha_{,RRR} \,.
  \label{D11a}
\end{equation}
Then, by specifying $\alpha(R)$ and providing a particular Hubble evolution $H(t)$, the function $\gamma(t)$ is determined. The complete function $\gamma(X,\phi)$ is provided by assuming a particular form of the scalar field Lagrangian that satisfy
\begin{equation}
  \gamma_{,X} X=\frac{1}{2}\alpha+\left(3H^2-\frac{1}{2}R\right)\alpha_{,R}+3H\dot{\alpha}_{,R}-\kappa^2\rho_m+\frac{1}{2}\gamma \,.
  \label{D11b}
\end{equation}
In order to reconstruct some particular solutions, we are considering the following $\alpha(R)$,
\begin{equation}
  \alpha(R)=R+aR^2 \,,
  \label{D12}
\end{equation}
where $a$ is a constant free parameter with the appropriate dimensions. This gravitational action is a very well known one in the literature as it is capable of reproducing inflation \cite{Starobinsky:1980te}. As above,  we start by considering a pure de Sitter solution (\ref{ds}), $H=H_0$. In such case,  the expression of  $\gamma(t)$ in (\ref{D11a}) yields
\begin{eqnarray}
  \gamma=-6H_0^2 \,.
\label{D13}
\end{eqnarray}
We can assume again $\gamma(X,\phi)=X-V(\phi)$ for illustrating the full reconstruction of the action (\ref{action2}),  and redefine the scalar field as $\phi=t$. Then, the kinetic term $X(\phi)$  and the scalar potential $V(\phi)$ turn out
\begin{equation}
  X(\phi)=X_0\e^{-3H_0 \phi} \,, \qquad V(\phi)=X_0\e^{-3H_0 \phi}+6H_0^2 \,,
\label{D14}
\end{equation}
where $X_0=-\kappa^2\rho_0$.

In the same way  as above, power-law solutions can be easily reconstructed. By assuming $H(t)=\frac{n}{t}$, the expression (\ref{D11a}) leads to
\begin{equation}
  \gamma(t)=2n\frac{18a (4-11n+6n^2)+(2-3n)t^2}{t^4} \,,
  \label{D15}
\end{equation}
while the kinetic term and the scalar potential yield
\begin{eqnarray}
  X(\phi)&=&X_0\phi^{-3n}+2n\frac{a(36-72n)+\phi^2}{\phi^4} \,, \nn
  V(\phi)&=& X_0\phi^{-3n}+\frac{-36an (2+n(-7+6n))+2n(-1+3n)\phi^2}{\phi^4} \,,
\label{D16}
\end{eqnarray}
where we have again assumed $\gamma(X,\phi)=X-V(\phi)$ for illustrative purposes.

Finally, let us reconstruct the gravitational action for $\Lambda$CDM model. In such a case, the Hubble parameter can be expressed as follows
\begin{equation}
  H(t)=\sqrt{\frac{\Lambda}{3}}\tanh\left[\frac{1}{2}\sqrt{3\Lambda}(t-t_0)\right] \,,
  \label{D17}
\end{equation}
where $\Lambda$ and $t_0$ are constants. Then, by using the equation (\ref{D11a}), the expression for $\gamma(t)$ yields
\begin{equation}
  \gamma(t)=-2\Lambda-9a\Lambda^2\sech^4\left[\frac{1}{2}\sqrt{3\Lambda}(t-t_0)\right] \,.
  \label{D18}
\end{equation}
While assuming $\gamma(X,\phi)=X-V(\phi)$, the kinetic term and the scalar potential result
\begin{eqnarray}
  X(\phi)&=&-9a\Lambda^2\sech^4\left[\frac{1}{2}\sqrt{3\Lambda}(\phi-t_0)\right] \,, \nn
  V(\phi)&=&2\Lambda \,.
\end{eqnarray}
Hence, the scalar potential reduces to a cosmological constant, while the dynamics of the scalar field compensates the extra terms in the gravitational sector (\ref{D12}).

Another interesting set of cosmological models are the so-called scaling solutions, which are characterized by
\begin{equation}
  \frac{\rho}{\rho_m} = C \,,
  \label{D19}
\end{equation}
where $C$ is a constant, and $\rho$ is the energy density of an additional perfect fluid. We can rewrite the first FLRW equation (\ref{D11}) as follows
\begin{equation}
  3\frac{\alpha_{,R}}{\kappa^2}H^2= \rho + \rho_m \,,
  \label{D20}
\end{equation}
where 
\begin{equation}
  \rho=\frac{1}{\kappa^2}\left[\gamma_{,X}X + \frac{1}{2}\left( \alpha_{,R} R-\alpha-\gamma \right) - 3H \dot{\alpha}_{,R} \right] \,,
  \label{D21}
\end{equation}
which leads to the constraint equation:
\begin{equation}
  \frac{1}{\kappa^2}\left[\gamma_{,X}X + \frac{1}{2}\left( \alpha_{,R} R-\alpha-\gamma \right) - 
  3H \dot{\alpha}_{,R}\right]-C\rho_{m0}\e^{-3\int dt H(t)} = 0 \,.
  \label{D22}
\end{equation}
By considering the Hilbert-Einstein action, $\alpha(R)=R$, together with a kind of k-essence field that satisfies $\rho/\rho_m=C$, the equation (\ref{D20}) can be easily solved by using the continuity equation leading to
\begin{equation}
  a(t)\propto t^{2/3} \qquad H(t)=\frac{2}{3t} \,,
  \label{D23}
\end{equation}
while the equation (\ref{D22}) leads to
\begin{equation}
  \gamma_{,X}X-\frac{1}{2}\gamma=C\frac{\kappa^2 \rho_0}{t^2} \,.
  \label{D24}
\end{equation}
And the second FLRW equation (\ref{D11}) provides an additional condition over the scalar field sector:
\begin{equation}
  \gamma_{,X}X=\frac{4-3\kappa^2\rho_0}{3t^2} \,,
  \label{D25}
\end{equation}
which together with (\ref{D24}) allows to reconstruct the scalar field Lagrangian once a particular form of $\gamma$ is assumed. In order to illustrate the procedure, let us consider the usual quintessence field $\gamma=X(\phi)-V(\phi)$. Then, the kinetic term and the scalar potential yield
\begin{eqnarray}
  X(\phi) & = & \frac{4-3\kappa^2\rho_0}{3\phi^2} \,, \nn
  V(\phi) & = & \frac{\kappa^2 \rho_0(6C+1)-4}{3\phi^2} \,,
  \label{D26}
\end{eqnarray}
where once again we have redefined the scalar field as the cosmic time, $\phi=t$. Note that the potential (\ref{D26}) is very well known as one of the possible scalar potentials that provides scaling solutions within quintessence models, others are the exponential ones $V(\phi)\propto\e^{\phi}$, see Refs.~\cite{Tsujikawa:2004dp,Copeland:2004qe}.

\section{Conclusions}
\label{conclusions}

In this work, we explored a generalised modified gravity theory, namely, $f(R,\phi,X)$ gravity, where $R$ is the Ricci scalar, $\phi$ a scalar field, and $X$ a kinetic term. This theory contains a wide range of dark energy and modified gravity models, particularly higher derivative models, both in the gravitational sector and also in the matter one. Some models within this action include $f(R)$ modified gravity and Galileons, which have been widely studied in the literature. We have then considered some specific Hubble parameters, specifically those which reproduce late-time cosmic acceleration and a realistic cosmological evolution, and the corresponding $f(R,\phi,X)$ action has been reconstructed.

Using the reconstruction technique studied here, the gravitational action can be easily obtained once some restrictions are assumed on the kind of function $f(R,\phi,X)$. We note that this action carries several extra degrees of freedom, so a particular cosmological solution does not provide a unique action. A degeneracy problem as in every dark energy model. However we suggest some simple actions that reproduce late-time acceleration and even the exact $\Lambda$CDM behaviour. Future works should be devoted to studying the viability of the different actions within $f(R,\phi,X)$ gravity framework. Besides the cosmological evolution, one should study cosmological perturbation theory with the view of understanding the growth of structure. Another interesting point may arise when analysing screening-like mechanisms, similar to the chameleon mechanism in this framework. Note also that the cases of the gravitational action (\ref{action1}) studied here, stand for a minority of the wide class of models included in the action (\ref{action1}), some interesting ones to be considered in future papers may be of the type  $f_1(R)f_2(X,\phi)$. Taking into account various observational facts about cosmology, it is hoped that one can restrict the form of the action $f(R,\phi,X)$ and provide a clearer way for explaining the late-time acceleration and even its unification with the inflationary paradigm.

\subsection*{Acknowledgements}

FSNL acknowledges financial support of the Funda\c{c}\~{a}o para a Ci\^{e}ncia e a Tecnologia (Portugal) through an Investigador FCT Research contract, with reference IF/00859/2012, and the grants EXPL/FIS-AST/1608/2013 and PEst-OE/FIS/UI2751/2014. DSG acknowledges support from a postdoctoral fellowship Ref.~SFRH/BPD/95939/2013 by the FCT (Portugal). FSNL and DSG were also supported by Funda\c{c}\~{a}o para a Ci\^{e}ncia e a Tecnologia (FCT) through the research grant UID/FIS/04434/2013. SB is supported by the Comisi\'on Nacional de Investigaci\'on Cient\'{\i}fica y Tecnol\'ogica (Becas Chile Grant No.~72150066).


\end{document}